\documentclass[aps,prl,showpacs,twocolumn,a4paper]{revtex4}
\usepackage{graphicx}
\usepackage{amsmath}
\usepackage{amsfonts}

\begin{document}

\title{Magnetic Flux Tuning of Spin Chirality in Mott Insulators with Ring Exchanges}
\author{Yi-Fei Wang$^{1}$ and Chang-De Gong$^{1,2}$}
\affiliation{$^1$Center for Statistical and Theoretical Condensed
Matter Physics, and Department of Physics, Zhejiang Normal
University, Jinhua 321004, China\\$^2$National Laboratory of Solid
State Microstructures and Department of Physics, Nanjing University,
Nanjing 210093, China}
\date{\today}

\begin{abstract}
A manifestation of the many-body Aharonov-Bohm effect in the
magnetic-flux-tuned Mott insulators with three-spin and four-spin
ring exchanges, presents as an effective tool to manipulate the
ground-state spin chirality, such as, tune the magnitude
continuously, switch an abrupt jump, or even reverse its sign. Such
a mechanism is demonstrated explicitly in both quasi-one-dimensional
ladders and two-dimensional lattices with triangles as elementary
plaquettes.
\end{abstract}

 \maketitle

{\it Introduction.---}Mott insulators (MIs), as the paradigm of
strongly correlated materials, have been commonly considered to have
only magnetic properties at low energies due to their spin moments.
However, it has been demonstrated recently by Bulaevskii {\it et
al.}~\cite{Bulaevskii} that, due to a certain form of charge
fluctuations, geometrically frustrated MIs may exhibit electric
orbital currents accompanying spin textures with chirality. This
notion of spin chirality itself has already been an intriguing topic
for decades in quantum magnetism, superconductivity and anomalous
Hall effect~\cite{Taguchi}. And the effective spin Hamiltonian of
MIs may contain a linear coupling of the spin chirality to an
external magnetic field, as proposed by Motrunich~\cite{Motrunich1}
when studying the organic compound $\kappa$-(ET)$_2$Cu$_2$(CN)$_3$
in which possible spin liquids with spinon Fermi surfaces are of
particular interest~\cite{Motrunich2,PALee,Motrunich3,Sheng1}.

At the heart of the theories by Bulaevskii {\it et al.} and
Motrunich~\cite{Bulaevskii,Motrunich1}, it is the three-spin ring
exchange (3SRE) in addition to the Heisenberg antiferromagnetic
(AFM) two-spin coupling, which is a specific form of charge
fluctuations in MIs~\cite{Takahashi,Rokhsar}. This multiple-spin
exchange concept, initiated by Thouless~\cite{Thouless}, now appears
as essential in various strongly-correlated systems: bcc solid
$^3$He~\cite{Roger1}, solid $^3$He films adsorbed on
graphite~\cite{Momoi,Misguich}, two-dimensional (2D) electron Wigner
crystals~\cite{Roger2,Bernu}, and zigzag Wigner crystals in quantum
wires~\cite{Meyer}. Several experiments have confirmed the presence
of four-spin ring exchange (4SRE) in cuprates~\cite{Coldea,Toader}.
And loading cold atoms into optical lattices opens another avenue to
design ring exchanges~\cite{Buchler}.

Despite the above encouraging advances, one still has no
quantitative understanding that how can we effectively manipulate
the spin chirality in MIs with ring exchanges. It is both of
interest and timely to address this problem, and here we conduct
such a study of a frustrated spin-$1/2$ system with 3SRE and 4SRE
modulated by a magnetic flux. Employing exact diagonalization (ED)
of finite systems, we consider both a (two-leg) triangular ladder
and a 2D triangular lattice geometry with periodic boundary
conditions (PBCs), which are the simplest systems on which both 3SRE
and 4SRE are possible. Beyond the weak-magnetic-flux regime, we
explore the large parameter space systematically, and demonstrate
that: at specific combinations of exchange interactions, varying the
magnetic flux strength enable us to tune continuously the magnitude,
switch an abrupt jump, or even change the sign, of the ground-state
(GS) spin chirality.

{\it Model Hamiltonian.---}With the nearest-neighbor (NN) Heisenberg
AFM coupling, the 3SRE and 4SRE terms modulated by a uniform
magnetic flux, the spin-$1/2$ model Hamiltonian in a triangular
ladder/lattice reads
\begin{eqnarray}\nonumber
H&=&J\sum_{\langle
ij\rangle}\mathbf{S}_{i}\cdot\mathbf{S}_{j}-K_{3}\sum_{ijk\in\triangle}\left[e^{i\phi}P_{ijk}+e^{-i\phi}P^{-1}_{ijk}\right]\\
&+&K_{4}
\sum_{ijkl\in\lozenge}\left[e^{i2\phi}P_{ijkl}+e^{-i2\phi}P^{-1}_{ijkl}\right]
\label{e.1}
\end{eqnarray}
where $\mathbf{S}_{i}$ is the spin operator on site $i$. $P_{ijk}$
which defined as
$P_{123}:|\sigma_1,\sigma_2,\sigma_3\rangle\rightarrow|\sigma_3,\sigma_1,\sigma_2\rangle$,
is the cyclic permutation of the three spins sitting on a triangular
plaquette, and satisfies $P^{-1}_{123}=P^{\dagger}_{123}=P_{321}$.
And similarly
$P_{1234}:|\sigma_1,\sigma_2,\sigma_3,\sigma_4\rangle\rightarrow|\sigma_4,\sigma_1,\sigma_2,\sigma_3\rangle$
is the cyclic permutation of the four spins sitting on a rhombus
consisting of two elementary triangles. The three sums in
Eq.~(\ref{e.1}) run, respectively, over all NN bonds, elementary
triangles and rhombi. Contrary to $^3$He systems in which the $^3$He
atoms are neutral, and similar to electron Wigner
crystals~\cite{Okamoto}, a magnetic flux through the exchange path
can change the nature of the ring exchanges in MIs, owing to the
Aharonov-Bohm (AB) effect. $\phi$ is the magnetic flux treading a
triangular plaquette, in units of $\phi_0/2\pi$ ($\phi_0=hc/e$ is
the flux quantum). We focus on the physical parameter space with
$J,K_{3}>0$ and $K_{4}\geq 0$, and vary the ratio $J/K_{3}$ and
$K_{4}/K_{3}$ with the setting $K_{3}=1$ (as an energy unit) in the
following calculations.

\begin{figure}[!htb]
  \vspace{-0.4in}
  \hspace{-0.2in}
  \includegraphics[scale=0.70]{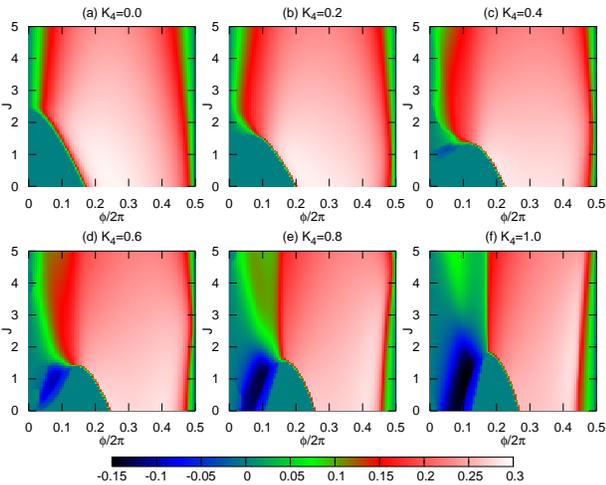}
  \vspace{-0.1in}
  \caption{(color online). Intensity plots of the GS spin chirality $\chi_{0}$ in the $\phi$-$J$ parameter space (setting $K_3=1$) of
  the $8\times 2$ triangular ladder at various fixed 4SRE strengths $K_{4}$'s.} \label{f.1}
\end{figure}

For our spin-$1/2$ case, $P_{ijk}$ can also be written in terms of
electron operators as $P_{123}=(c^{\dagger}_{1\alpha}c_{1\beta})
(c^{\dagger}_{2\beta}c_{2\gamma})
(c^{\dagger}_{3\gamma}c_{3\alpha})$, and satisfies
$i(P_{123}-P_{321})=-4\mathbf{S}_{1}\cdot \mathbf{S}_{2}\times
\mathbf{S}_{3}\equiv -4\chi_{123}$, where $\chi_{ijk}$ represents
the local spin chirality~\cite{Taguchi}. In previous studies where a
magnetic flux is absent, the 3SRE term itself favors ferromagnetism
(FM) and can be taken into account just by modifying $J$ and
allowing both $J>0$ and $J\leq0$~\cite{Momoi,Misguich}. However even
for a small $\phi$, since $e^{i\phi}P_{123}+e^{-i\phi}P_{321}
=\cos\phi(P_{123}+P_{321}) -4\sin\phi(\mathbf{S}_{1}\cdot
\mathbf{S}_{2}\times \mathbf{S}_{3})\simeq(P_{123}+P_{321})
-4\phi\chi_{123}$, the magnetic flux couples linearly to the spin
chirality in the low-$\phi$
limit~\cite{Bulaevskii,Motrunich1,Rokhsar}, and therefore could
probably induce a non-zero chirality density. Here we are concerned
with the parameter space of $\phi\in[0,\pi]$, and hence consider the
many-body AB effect induced by a strong magnetic flux.

{\it Triangular ladders.---}We firstly consider the ladder geometry.
We study the GS averaged (local) spin chirality
$\chi_{0}\equiv-\langle\chi_{ijk}\rangle_0$ (here a negative sign is
added for convenience) by varying $\phi$, $J$ and $K_{4}$. The
typical ED results for a triangular ladder of the size $8\times 2$
are shown in Fig.~\ref{f.1}.

In the absence of 4SRE ($K_{4}=0$), $\chi_{0}$ is non-negative in
the parameter region $\phi\in[0,\pi]$, as shown in
Fig.~\ref{f.1}(a). Note that $\chi_{0}$ has the symmetry
$\chi_{0}(\phi)=-\chi_{0}(-\phi)=-\chi_{0}(2\pi-\phi)$, which has
been numerically confirmed. The whole $\phi$-$J$ parameter space is
roughly separated into two regions: the bottom left corner
($J\lesssim2.4$, $\phi/2\pi\lesssim0.17$, and uniformly colored)
with $\chi_0=0$ and saturated FM ($S_{\rm{tot}}=S_{\rm{max}}$), and
larger region with $\chi_0>0$ and spin-singlet GSs
($S_{\rm{tot}}=0$). The quantum critical line between these two
regions has also been verified through tracking the
non-analyticities in the GS energy function $E_{0}(\phi,J)$. Later,
we will show that the quantum critical line does not depend
appreciably on the ladder size.

For the triangular ladders with PBCs, because of the identity
$P_{123}+P_{321}=2\mathbf{S}_{1}\cdot\mathbf{S}_{2}
+2\mathbf{S}_{2}\cdot\mathbf{S}_{3}
+2\mathbf{S}_{3}\cdot\mathbf{S}_{1}+1/2$, the Hamiltonian with
$K_{4}=0$ will reduce to $H=-N\cos\phi+\widetilde{J}_{1}\sum^{\rm
inter}_{\langle
ij\rangle}\mathbf{S}_{i}\cdot\mathbf{S}_{j}+\widetilde{J}_{2}\sum^{\rm
intra}_{\langle ij\rangle}\mathbf{S}_{i}\cdot\mathbf{S}_{j}
+4\sin{\phi}\sum_{ijk\in\triangle}\mathbf{S}_{i}\cdot
\mathbf{S}_{j}\times \mathbf{S}_{k}$ with
$\widetilde{J}_{1}=J-4\cos\phi$ and $\widetilde{J}_{2}=J-2\cos\phi$,
where the superscript ``inter'' (``intra'') corresponds to the
effective interchain (intrachain) two-spin coupling
$\widetilde{J}_{1}$ ($\widetilde{J}_{2}$). At the left boundary line
of Fig.~\ref{f.1}(a) with $\phi=0$, there is a quantum critical
point $J\approx2.4$ corresponding to
$\widetilde{J}_{2}/\widetilde{J}_{1}=-0.25$ which separates the
saturated FM phase~\cite{Hamada} and the dimer
phase~\cite{Majumdar}. And at the right boundary line of
Fig.~\ref{f.1}(a) with $\phi=\pi$, the GS is also the dimer phase
since $J\geq0$ and $\phi=\pi$ gives
$\widetilde{J}_{2}/\widetilde{J}_{1}\geq0.5$~\cite{Tonegawa}. The
features of these states will be displayed and discussed later by
various correlation functions.

In the presence of 4SREs, there are even more interesting behaviors
of $\chi_0$, as shown in Figs.~\ref{f.1}(b)-(f) with five typical
$K_{4}$'s respectively. At $K_{4}=0.2$ [Fig.~\ref{f.1}(b)], the
saturated FM region shrinks in the $J$ direction while expand a
little in the $\phi$ direction. At $K_{4}=0.4$ and $K_{4}=0.6$
[Figs.~\ref{f.1}(c) and (d)], at the center of saturated FM region,
there appears a negative-$\chi_0$ region in which the GSs are spin
singlets ($S_{\rm{tot}}=0$). When $K_{4}$ is further increased to
$0.8$ [Fig.~\ref{f.1}(e)] and $1.0$ [Fig.~\ref{f.1}(f)], the
negative-$\chi_0$ region continues to expand and occupies a
significant portion in the $\phi$-$J$ parameter space.

\begin{figure}[!htb]
  \vspace{-0.0in}
  \hspace{-0.15in}
  \includegraphics[scale=0.6]{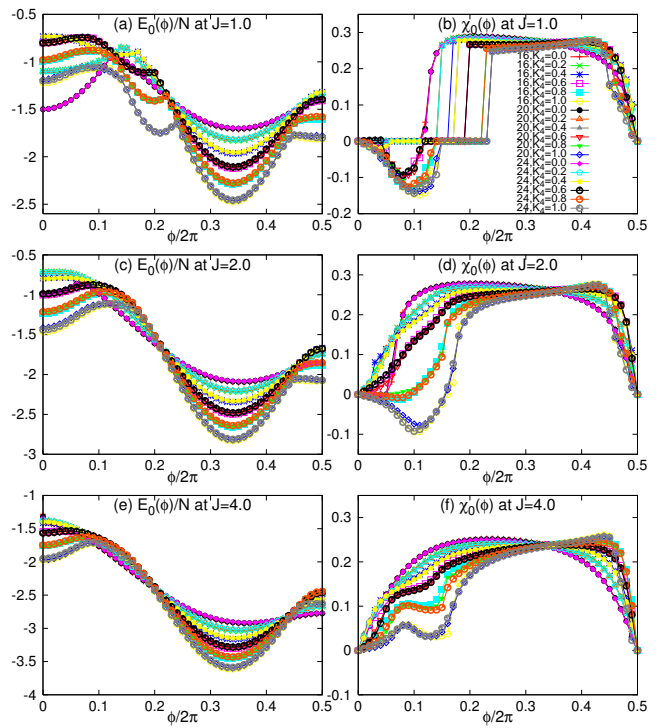}
  \vspace{-0.2in}
  \caption{(color online). Triangular ladders: GS energy per site $E_{0}/N$ (left) and GS spin chirality $\chi_{0}$ (right) versus $\phi$,
  for various $J$'s, $K_4$'s and ladder sizes $N$'s.} \label{f.2}
\end{figure}

An intuitive analysis for $K_{4}>0$ is much more difficult, than
that in the simpler case of $K_{4}=0$. However, it should be noted
that the 4SRE operators satisfy
$P_{1234}-P_{4321}={1\over{2}}(P_{123}+P_{234}+P_{341}+P_{412}-{\rm
H.c.})=2i(\chi_{123}+\chi_{234}+\chi_{341}+\chi_{412})$. Therefore,
$e^{i2\phi}P_{1234}+e^{-i2\phi}P_{4321}
=\cos2\phi(P_{1234}+P_{4321})-2\sin2\phi(\chi_{123}+\chi_{234}+\chi_{341}+\chi_{412})$.
Due to the opposite signs and the different AB periods of 3SRE and
4SRE terms, the low-$\phi$-limit coupling coefficient and the
portion of negative-$\chi_0$ region depend on the competitions
between them.

In order to address the effects of ladder sizes, we compare three
sizes of $8\times 2$, $10\times 2$ and $12\times 2$. The mainly
considered quantities are $\chi_0$ and the GS energy per site
$E_0/N$. From Figs.~\ref{f.2}(a),(c),(e), we can see that the
$E_0(\phi)/N$ curves coincide well with each other for all three
ladder sizes. And the $\chi_0(\phi)$ curves
[Figs.~\ref{f.2}(b),(d),(f)] also present us rather identical
behaviors. All these results indicate that in the thermodynamic
limit ($N\rightarrow \infty$), both $E_0(\phi)/N$ and $\chi_0(\phi)$
will not deviate obviously from these finite-size results.

\begin{figure}[!htb]
  \vspace{-0.0in}
  \hspace{-0.15in}
  \includegraphics[scale=0.6]{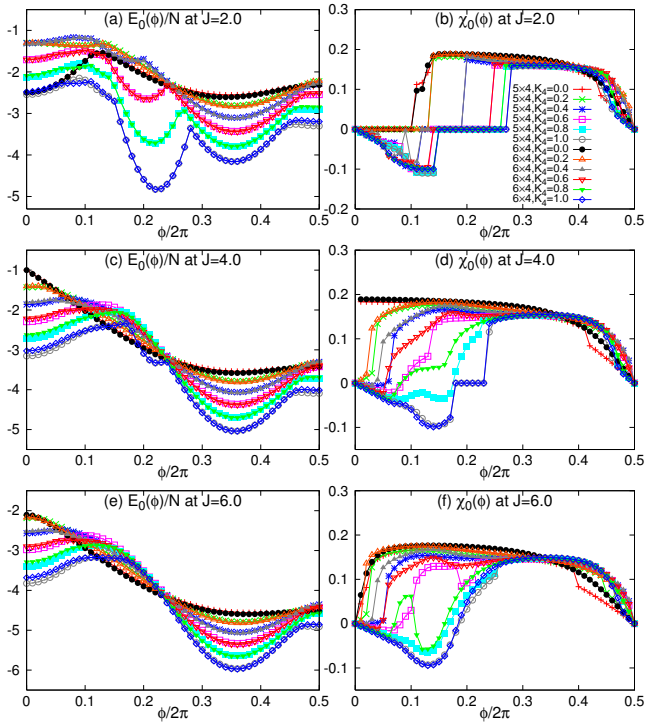}
  \vspace{-0.2in}
  \caption{(color online). Triangular lattices: $E_{0}/N$ (left) and $\chi_{0}$ (right) versus $\phi$,
  for various $J$'s, $K_4$'s and ladder sizes $N$'s.} \label{f.3}
\end{figure}

{\it Triangular lattices.---}For the 2D triangular lattices, we
focus on two cases with the sizes of $5\times 4$ and $6\times 4$
[Fig.~\ref{f.3}]. In the absence of 4SRE terms ($K_{4}=0$), similar
to the previous case of ladders, the Hamiltonian of a 2D triangular
lattice with PBCs will reduce to
$H=-N\cos\phi+(J-4\cos\phi)\sum_{\langle
ij\rangle}\mathbf{S}_{i}\cdot\mathbf{S}_{j}+4\sin{\phi}\sum_{ijk\in\triangle}\mathbf{S}_{i}\cdot
\mathbf{S}_{j}\times \mathbf{S}_{k}$. From Fig.~\ref{f.3}, we can
see that $E_0(\phi)/N$ and $\chi_0(\phi)$ display quite similar
behaviors resulting from the competitions between 3SRE and 4SRE
terms as
the ladders, such as the sign changes and abrupt jumps of $\chi_0$. 

{\it Long-range correlations and ordering.---}We now turn to three
kinds of correlation functions (CFs) in ladders. The first is the
spin-spin CF, defined as $C(r)=\langle \mathbf{S}_{i}\cdot
\mathbf{S}_{i+r}\rangle$, where $r$ is the range (in units of the
lattice constant) between two sites along the chain direction and
takes integer (half-integer) values for intrachain (interchain)
spin-spin CF $C_{1}(r)$ [$C_{2}(r)$]. The other two CFs are defined
as follows~\cite{Sheng1,Misguich}. The dimer operator on a bond
$(i,j)$ is defined by $d_{ij}=(1-P_{ij})/2$ (where $P_{ij}$
exchanges two spins as
$P_{12}:|\sigma_1,\sigma_2\rangle\rightarrow|\sigma_2,\sigma_1\rangle$),
and this projector gives $1$ on a singlet and $0$ on a triplet. The
dimer-dimer CF between two bonds is $D(r)=\langle
d_{ij}d_{kl}\rangle-\langle d_{ij}\rangle \langle d_{kl}\rangle$,
and $D_{1}(r)$ [$D_{2}(r)$] for two parallel (non-parallel) rung
bonds between two chains. The chiral-chiral CF between two triangles
is defined as $X(r)=\langle \chi_{ijk}\chi_{lmn}\rangle$, and
$X_{1}(r)$ for two up-triangles (or equivalently two
down-triangles),  while $X_{2}(r)$ for an up-triangle  and a
down-triangle. Note that if the two triangles have some sites in
common, $X(r)$ may have a small imaginary part, and only the real
part is plotted~\cite{Misguich}.

At each case when $S_{\text{tot}}=S_{\text{max}}=N/2$, $\chi_0$
always vanishes, and meanwhile $C(r)$ is surely positive and almost
a constant at any range $r$. Now we would take a closer look at the
three kinds of CFs of the other cases, and focus on the $12\times2$
triangular ladder as an example.

\begin{figure}[!htb]
  \vspace{-0.1in}
  \hspace{-0.2in}
  \includegraphics[scale=0.63]{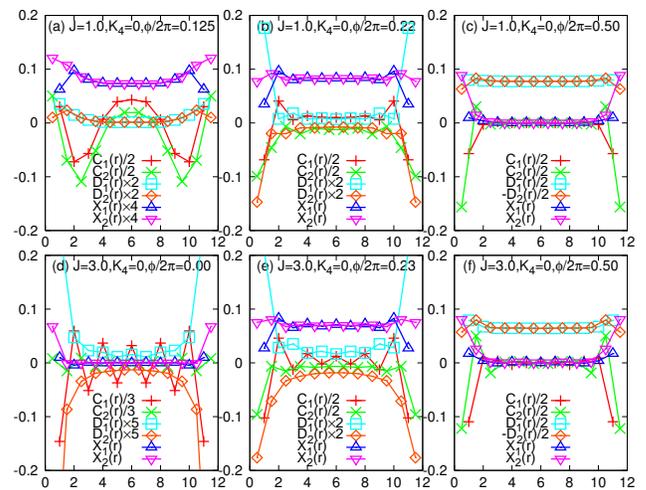}
  \vspace{-0.2in}
  \caption{(color online). Various GS correlation functions (see text) versus the range $r$ of the $12\times2$ triangular ladders at various $J$'s, $\phi$'s and $K_{4}=0$.
} \label{f.4}
\end{figure}

We first consider the simpler cases with only 3SREs (Fig.~\ref{f.4}
with $K_{4}=0$). For $J=1.0$, tuning $\phi/2\pi$ across an FM
quantum critical point at $0.125$ [Fig.~\ref{f.4}(a)], both
$C_{1}(r)$ and $C_{2}(r)$ exhibit that the GS consists of two-period
FM domains with opposite magnetization, which is a remnant signature
of long-range FM ordering in the FM region; tuning $\phi/2\pi$
further to $0.22$ [Fig.~\ref{f.4}(b)] at which $\chi_0(\phi)$ takes
a maximum, $C_{1}(r)$ [$C_{2}(r)$] shows weak FM (AFM) correlations,
and both $X_{1}(r)$ and $X_{2}(r)$ reveal nondecaying long-range
correlations; when $\phi$ is increased to $\pi$ [Fig.~\ref{f.4}(c)],
$C(r)$'s and $X(r)$'s show fast decaying behaviors, while $D(r)$'s
reveal the long-range dimer ordering. For $J=3.0$, $\phi=0$
[Fig.~\ref{f.4}(d)], $C_{1}(r)$ [$C_{2}(r)$] shows strong (weak) AFM
correlations because of $\widetilde{J}_{2}>\widetilde{J}_{1}$, and
$D(r)$'s also show slowly decaying correlations; tuning $\phi/2\pi$
to make $\chi_0(\phi)$ take a maximum [Fig.~\ref{f.4}(e)] and then
to $0.5$ [Fig.~\ref{f.4}(f)], there kinds of CFs resemble the
$J=1.0$ cases.

\begin{figure}[!htb]
  \vspace{-0.1in}
  \hspace{-0.2in}
  \includegraphics[scale=0.63]{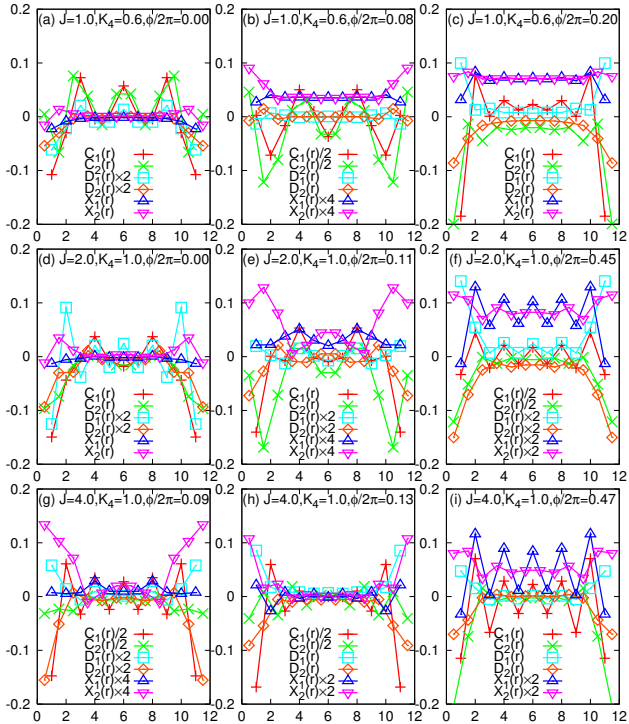}
  \vspace{-0.2in}
  \caption{(color online). The same as Fig.~\ref{f.4}, but with
  nonzero 4SREs ($K_4>0$).
   } \label{f.5}
\end{figure}

Next, we progress to much more complicated and interesting cases
with 4SREs (Fig.~\ref{f.5}). For $(J,K_{4})=(1.0,0.6)$, at $\phi=0$
[Fig.~\ref{f.5}(a)], the nonzero $K_{4}$ makes the long-range FM
correlations destroyed, $C(r)$'s and $D(r)$'s exhibit nondecaying
fluctuations although $X(r)$'s show fast decaying behaviors; tuning
$\phi/2\pi$ to $0.08$ [Fig.~\ref{f.5}(b)], at which $\chi_0(\phi)$
takes a negative minimum, the $C(r)$'s exhibit FM domains and
$X(r)$'s reveal long-range correlations; then tuning $\phi/2\pi$ to
$0.20$ [Fig.~\ref{f.5}(c)], at which $\chi_0(\phi)$ jumps to a large
positive value, $X(r)$'s show long-range correlations. For
$(J,K_{4})=(2.0,1.0)$, tuning $\phi/2\pi$ to $0.11$
[Fig.~\ref{f.5}(e)] at which $\chi_0(\phi)$ takes a minimum, the
$X(r)$'s show strong two-period fluctuations; tuning $\phi/2\pi$
further to $0.45$ [Fig.~\ref{f.5}(f)] at which $\chi_0(\phi)$ drops
steeply, $X_{1}(r)$ displays strongly fluctuating correlations, and
$C_{1}(r)$ also shows strong intrachain AFM correlations. For
$(J,K_{4})=(4.0,1.0)$, tuning $\phi$ to $0.09$ [Fig.~\ref{f.5}(g)]
at which $\chi_0(\phi)$ takes a local maximum, the $X(r)$'s show
strong two-period fluctuations; when tuning $\phi/2\pi$ to $0.47$
[Fig.~\ref{f.5}(i)], we can see some behaviors resembling the
previous case in Fig.~\ref{f.5}(f).

{\it Summary and discussion.---}For a spin-$1/2$ system in a
triangular ladder/lattice with NN AFM coupling, 3SRE and 4SRE, and a
uniform magnetic flux $\phi$, we can effectively manipulate the GS
spin chirality $\chi_0$, such as tune continuously the magnitude of
$\chi_0$ by varying $\phi$, switch an abrupt jump near an FM phase
boundary, or even reverse the sign of $\chi_0$, and change the
low-$\phi$-limit coupling coefficient. Various CFs discover the
characteristic long-range correlations accompanying the tuned spin
chiralities. Such a mechanism presents a peculiar manifestation of
the many-body AB effect on quasi-localized spins in MIs. This
magnetic flux tuning of spin chirality is expected be observed in 2D
organic compound $\kappa$-(ET)$_2$Cu$_2$(CN)$_3$, quasi-1D and 2D
Wigner crystals, and cold atoms in optical lattices with ring
exchanges.

This work was supported by NSFC of China (No. 10904130) and State
Key Program for Basic Researches of China (No. 2006CB921802). ED
calculations were based on TITPACK Ver. 2 package by H. Nishimori.

\end{document}